\documentclass[fleqn,12pt,twoside]{article}
\usepackage{espcrc1}

\usepackage{graphicx}
\usepackage[figuresright]{rotating}
\usepackage{epsfig}  

\newcommand\cO{\mathcal{O}} 

\newcommand{\AmS}{{\protect\the\textfont2
  A\kern-.1667em\lower.5ex\hbox{M}\kern-.125emS}}
\newcommand{\Dd}[1]{\mbox{
  \parbox[b]{0cm}{$D$}\raisebox{1.7ex}{$\leftrightarrow$}$_{\!#1}$}}

\title{\vspace{-3.65cm}
       {\normalsize LU-ITP 2002/015}    \\[-0.2cm]
       {\normalsize Edinburgh 2002/06}   \\
       \vspace{2.72cm}
Applied lattice gauge calculations: diquark content of the 
nucleon\thanks{Invited talk given by M. G\"ockeler at the European 
Workshop on the QCD Structure of the Nucleon
(QCD - N'02), Ferrara, Italy, 3-6 Apr 2002}}

\author{M. G\"ockeler\address[LEI]{Institut f\"ur Theoretische Physik,
           Universit\"at Leipzig, D-04109 Leipzig, Germany}$^,$%
        \address[REG]{Institut f\"ur Theoretische Physik,
           Universit\"at Regensburg, D-93040 Regensburg, Germany},
        R. Horsley\address{Dept.\ of Physics and Astronomy,
           University of Edinburgh, Edinburgh EH9 3JZ, UK},
        B. Klaus\address[NIC]{John von Neumann-Institut f\"ur
           Computing NIC, DESY, D-15735 Zeuthen, Germany}$^,$%
           \address[FUB]{Institut f\"ur 
           Theoretische Physik, Freie Universit\"at Berlin,
           D-14195 Berlin, Germany},
	D. Pleiter\addressmark[NIC],
	P.E.L.~Rakow\addressmark[REG],
	S. Schaefer\addressmark[REG],
	A.~Sch\"afer\addressmark[REG]
        and
        G. Schierholz\addressmark[NIC]$^,$\address{Deutsches 
	   Elektronen-Synchrotron DESY, D-22603 Hamburg, Germany}}
       
\begin{document}

\maketitle

\begin{abstract}
As an example of an application of lattice QCD we describe a computation 
of four-quark operators in the nucleon. The results are interpreted 
in a diquark language.
\end{abstract}

\section{INTRODUCTION}

Lattice gauge theory offers the possibility
to perform non-perturbative computations in a strongly coupled theory
without any model assumptions. Lattice QCD, in particular, allows us 
to investigate low-energy properties of QCD, like, e.g., the hadron mass
spectrum, decay constants, form factors, and moments of structure functions.
In this article, we shall discuss as an example the computation of 
four-quark operators in the nucleon recently performed by our group
(the QCDSF collaboration). Since four-quark operators are related to
higher-twist effects in nucleon structure functions, a subject of
considerable theoretical and experimental importance, they are a 
particularly interesting object for lattice-QCD studies. 
For more technical details we refer to
Ref.\cite{nucl4}. 

\section{WHAT CAN WE COMPUTE ON THE LATTICE?}

The basic observables in lattice QCD are Euclidean $n$-point correlation
functions. Since space-time has been discretised (with lattice spacing $a$)
the path integral has become a high-dimensional ordinary integral 
over a discrete set of field variables, which can be evaluated by
Monte Carlo methods. As the (Grassmann valued) quark 
fields appear bilinearly in the action, they can be integrated out 
analytically leaving behind the determinant of the lattice Dirac operator
and products of quark propagators. In the quenched approximation, which
will be employed throughout the whole paper, this determinant is replaced 
by 1. The quenched approximation saves a lot of computer time, but it
is hardly possible to estimate its accuracy.

Let us briefly sketch the computation of hadronic matrix elements.
First, one has to choose suitable interpolating fields
for the particle to be studied. For a proton with momentum 
$\vec{P}$ one may take
\begin{equation} 
  B_\alpha (t,\vec{P})  =  
\sum_{x;x_4=t}
\mathrm e^{- \mathrm i \vec{P}\cdot \vec{x} }
\epsilon_{i j k} u^i_\alpha (x) 
 u^j_\beta (x) (C \gamma_5)_{\beta \gamma} d^k_\gamma (x) 
\end{equation} 
and the corresponding $\bar{B}$, where $i$, $j$, \ldots are colour 
indices and $\alpha$, $\beta$, \ldots are Dirac indices.

As the time extent $T$ of our lattice tends to $\infty$, the
two-point correlation function becomes the vacuum expectation value of the
corresponding Hilbert space operators with the Euclidean evolution operator
${\mathrm e}^{- H t}$ in between, i.e.\ we have, omitting Dirac indices
and momenta for simplicity:
\begin{equation}
  \langle B(t) \bar{B}(0) \rangle 
   \stackrel{T \to \infty}{=} 
  \langle 0 |  B {\mathrm e}^{- H t}  \bar{B} | 0 \rangle \,.
\end{equation}
If in addition the time $t$ gets large, the ground state $|p \rangle$ of
the proton will dominate the sum over intermediate states between $B$
and $\bar{B}$, and the two-point function will decay exponentially with 
a decay rate given by the proton energy $E_p$:
\begin{equation}
  \langle B(t) \bar{B}(0) \rangle 
   \stackrel{T \to \infty}{=} 
  \langle 0 |  B {\mathrm e}^{- H t}  \bar{B} | 0 \rangle
   \stackrel{t \to \infty}{=} 
       \langle 0 |  B | p \rangle {\mathrm e}^{- E_p t}
       \langle p | \bar{B} | 0 \rangle + \cdots
\end{equation}
Of course, if the momentum vanishes, we have $E_p = m_p$, the proton mass.

Similarly we have for a three-point function with the operator $\cO$ whose
matrix elements we want to calculate: 
\begin{eqnarray}
  \langle B(t) \cO (\tau) \bar{B}(0) \rangle 
  & \stackrel{T \to \infty}{=} & 
    \langle 0 |  B {\mathrm e}^{- H (t-\tau)} \cO
       {\mathrm e}^{- H \tau} \bar{B} | 0 \rangle \nonumber
\\ 
  {} & = & \langle 0 |  B | p \rangle {\mathrm e}^{- E_p (t - \tau) }
      \langle p | \cO| p \rangle 
       {\mathrm e}^{-E_p \tau} \langle p | \bar{B} | 0 \rangle
      + \cdots \nonumber
\\
  {} & = & \langle 0 |  B | p \rangle {\mathrm e}^{- E_p t }
       \langle p | \bar{B} | 0 \rangle
       \langle p | \cO| p \rangle 
      + \cdots
\end{eqnarray}
if $t > \tau > 0$. Hence the ratio
\begin{equation} 
 R \equiv  
 \frac{ \langle B(t) \cO (\tau) \bar{B}(0) \rangle}
      { \langle B(t) \bar{B}(0) \rangle}
  = \langle p | \cO| p \rangle + \cdots 
\end{equation}
should be independent of the times $\tau$ and $t$, if all time differences
are so large that excited states can be neglected.

The proton three-point function for a two-quark operator contains quark-line
connected as well as quark-line disconnected pieces. In the quark-line
connected contributions the operator is inserted in one of the quark lines
of the nucleon propagator, while in the disconnected pieces 
the operator is attached to an additional closed quark line which  
communicates with the valence quarks in the proton only via gluon exchange.
In the limit of exact isospin invariance as it is considered in this paper,
the disconnected contributions cancel in the case of non-singlet 
two-quark operators. Moreover, one might try to argue that it would be 
more consistent with the quenched approximation to drop quark-line 
disconnected contributions generally.

\section{WHAT DO WE WANT TO COMPUTE ON THE LATTICE?}

The operator product expansion (OPE) relates moments of structure functions
to nucleon matrix elements of local operators. In the deep-inelastic limit,
where the momentum transfer $Q^2$ becomes large, one can express 
Nachtmann moments of, e.g., $F_2$ in the following form:
\begin{equation} 
\int_0^1 dx \, x^{n-2} F_2(x,Q^2)  \big|_{\mathrm {Nachtmann}}
 =  c^{(2)}_n A^{(2)}_n (\mu)
 + \frac{c^{(4)}_n}{Q^2} A^{(4)}_n (\mu)
 + O \left(\frac{1}{Q^4}\right) \,.
\label{nachtmann}
\end{equation}
Here $n=2,4,6,\ldots$, the (reduced) matrix elements $A^{(t)}_n$
of twist $t$ and spin $n$ are renormalised at the scale $\mu$, and 
$c^{(t)}_n = c^{(t)}_n(Q^2/\mu^2,g(\mu))$ are the Wilson coefficients,
calculated in perturbation theory. Note that we go back to Minkowski space
in this section.

Besides the leading twist-2 contribution we have included in 
(\ref{nachtmann}) the twist-4 
corrections. These are suppressed by $1/Q^2$ and have been the subject of
intensive investigations both experimentally and theoretically. In 
the following we shall describe what lattice computations can tell us about
twist-4 effects.

An important class of twist-4 operators consists of four-quark operators.
In particular, the twist-4, spin-2 matrix element $A_2^{(4)}$ is
defined by the following expectation value in a proton state with 4-momentum 
$P$ ($\{\cdots\}$ indicates symmetrisation)
\begin{equation} 
  \langle P | A^c_{\{\mu \nu \}} 
     - \mbox{traces} | P \rangle 
 = 2 A^{(4)}_2 (P_\mu P_\nu -  \mbox{traces} ) 
\end{equation}
with the four-quark operator
\begin{equation}
 A^c_{\mu \nu } = \bar{\psi} G \gamma_\mu \gamma_5 t^a \psi 
     \bar{\psi} G \gamma_\nu \gamma_5 t^a \psi \,.
\label{4qop}
\end{equation}
Here the quark field $\psi$ carries flavour, colour, and Dirac indices, the
$t^a$ are the usual generators of colour SU(3), and $G$ is a matrix
in flavour space, the charge matrix
\begin{equation} 
  G = \mbox{diag}(e_u,e_d) = \mbox{diag}(2/3,-1/3) \,.
\end{equation}

The corresponding leading-twist matrix element $A_2^{(2)}$
is defined by 
\begin{equation} 
  \langle P | \cO _{\{\mu \nu \}} 
     - \mbox{traces} | P \rangle 
 = 2 A^{(2)}_2 (P_\mu P_\nu -  \mbox{traces} ) 
\end{equation}
in terms of the two-quark operator 
\begin{equation}
  \cO _{\mu \nu} =   
       \frac{\mathrm i}{2} \bar{\psi} G^2 \gamma_\mu \Dd{\nu} \psi \,.
\label{2qop}
\end{equation}
This operator has dimension 4, while the four-quark operator (\ref{4qop})
has dimension 6. Hence four-quark operators may mix with operators of lower 
dimension (two-quark operators). This is a rather general phenomenon for 
operators of higher twist, because for given spin, higher twist means higher 
dimension. Unfortunately, the mixing with operators of lower dimension 
is hard to deal with in perturbation theory. For the time being we
do not attempt a non-perturbative calculation of the renormalisation
and mixing coefficients. Instead we look for four-quark operators for 
which mixing with two-quark operators is prohibited by flavour symmetry 
and apply perturbative (one-loop) renormalisation.

\section{FOUR-QUARK OPERATORS IN THE PROTON}

Assuming SU(2) flavour symmetry (isospin symmetry) we find that
two-quark operators can have isospin $I=0$ or 1, whereas four-quark operators
can have $I=0$, 1, or 2. Hence four-quark operators with $I=2$ cannot
mix with two-quark operators. Unfortunately, the expectation value
of any $I=2$ operator in the proton vanishes. So we have to enlarge 
the flavour symmetry from SU(2)$_{\mathrm F}$ to SU(3)$_{\mathrm F}$, 
i.e.\ we assume three quarks of the same mass, and
the flavour structure of the operator in the OPE is now
\begin{equation}
 \cO = (e_u \bar{u} u + e_d \bar{d} d + e_s \bar{s} s)
          (e_u \bar{u} u + e_d \bar{d} d + e_s \bar{s} s) \,.
\end{equation}
Whereas two-quark operators transform under SU(3)$_{\mathrm F}$ according
to $\overline{\mathbf{3}} \otimes \mathbf{3} = \mathbf{1} \oplus \mathbf{8}$,
we have for four-quark operators: $(\overline{\mathbf{3}} \otimes \mathbf{3})
\otimes (\overline{\mathbf{3}} \otimes \mathbf{3}) 
= 2 \cdot \mathbf{1} \oplus 4 \cdot \mathbf{8} \oplus
    \mathbf{10} \oplus \overline{\mathbf{10}} \oplus \mathbf{27}$ .
Four-quark operators with $I=0,1$, $I_3 = 0$, and hypercharge $Y=0$ from the 
$\mathbf{10}$, $\overline{\mathbf{10}}$, $\mathbf{27}$
multiplets do not mix with two-quark operators and can be used in a 
proton expectation value, e.g.\ the $I=1$ operator
\begin{eqnarray}
\cO^{I=1}_{27} & = &  \frac{1}{10} [
 (\bar{u} u) (\bar{u} u) - (\bar{d} d) (\bar{d} d) 
 - (\bar{u} s) (\bar{s} u) - (\bar{s} u) (\bar{u} s) 
 + (\bar{d} s) (\bar{s} d) + (\bar{s} d) (\bar{d} s)  \nonumber
\\ {} & {} &
 - (\bar{s} s) (\bar{u} u) - (\bar{u} u) (\bar{s} s) 
 + (\bar{s} s) (\bar{d} d) + (\bar{d} d) (\bar{s} s) ]
\end{eqnarray}
belongs to the $\mathbf{27}$ multiplet. Such a flavour component can
be projected out by suitable linear combinations of the matrix elements of
the octet baryons $p$, $n$, $\Lambda$, $\Sigma$, $\Xi$, e.g.
\begin{equation} 
 \langle \Sigma^+ | \cO | \Sigma^+ \rangle 
 -2 \langle \Sigma^0 | \cO | \Sigma^0 \rangle 
  + \langle \Sigma^- | \cO | \Sigma^- \rangle 
 = \langle p | \cO^{I=1}_{27} | p \rangle \,.
\end{equation}
Note that the flavour combination in four-quark operators which do not
mix with two-quark operators (like $\cO^{I=1}_{27}$) is such that all
quark-line disconnected contributions cancel. However, mixing with
other four-quark operators (with different Dirac or colour structures) 
remains and is taken into account using perturbatively calculated mixing
coefficients.  

After a chiral extrapolation (linear in the quark mass) we obtain in 
our special flavour channel
\begin{equation}
 \int_0^1 \mathrm dx \, F_2(x,Q^2) \big|_{\mathrm {Nachtmann}}
 ^{\mathbf{27}, I=1} = - 0.0005(5)
\frac{m_p^2 \alpha_s (Q^2)}{Q^2} + O(\alpha_s^2) \,.
\label{result}
\end{equation}
The $O(\alpha_s^2)$ contribution arises from the fact that we know only
the tree level Wilson coefficient $c_2^{(4)} = g^2 \left( 1+O(g^2) \right)$
\cite{wilco}.
For reasonable values of $Q^2$ the result is rather small using, e.g.,
$\alpha_s(Q^2) \approx 0.3$ for $Q^2=4 \, \mbox{GeV}^2$. But it is
interesting to note that the bag model estimates the prefactor in 
(\ref{result}) to be $\propto B/m_p^4 \approx 0.0006$, where 
$B \approx (145 \, \mbox{MeV})^4$ is the bag constant \cite{bag}.

\begin{figure}[htb]
\begin{center}
\vspace*{-1.0cm}
\epsfig{file=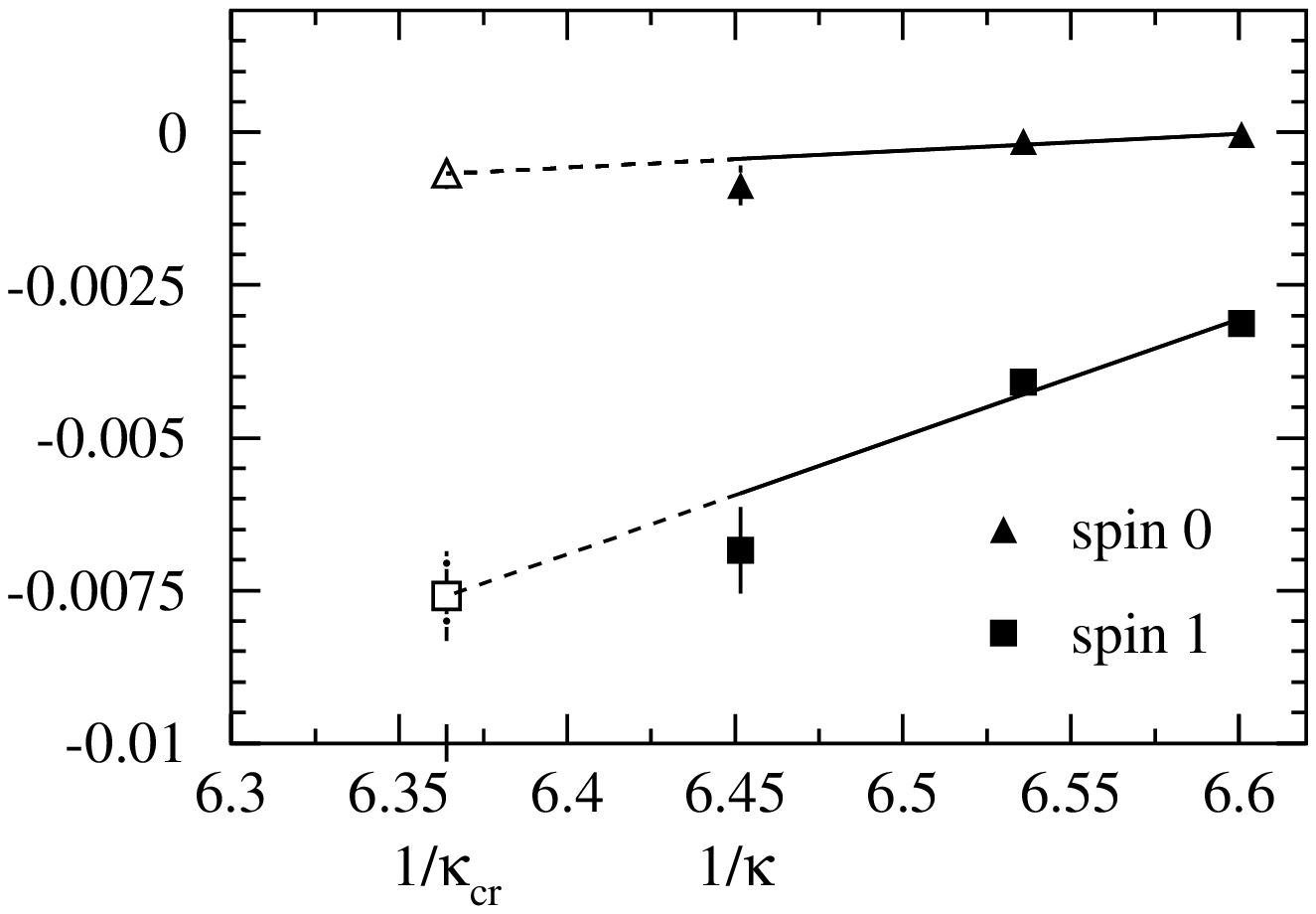,height=8cm}
\\[-1.0cm]
\epsfig{file=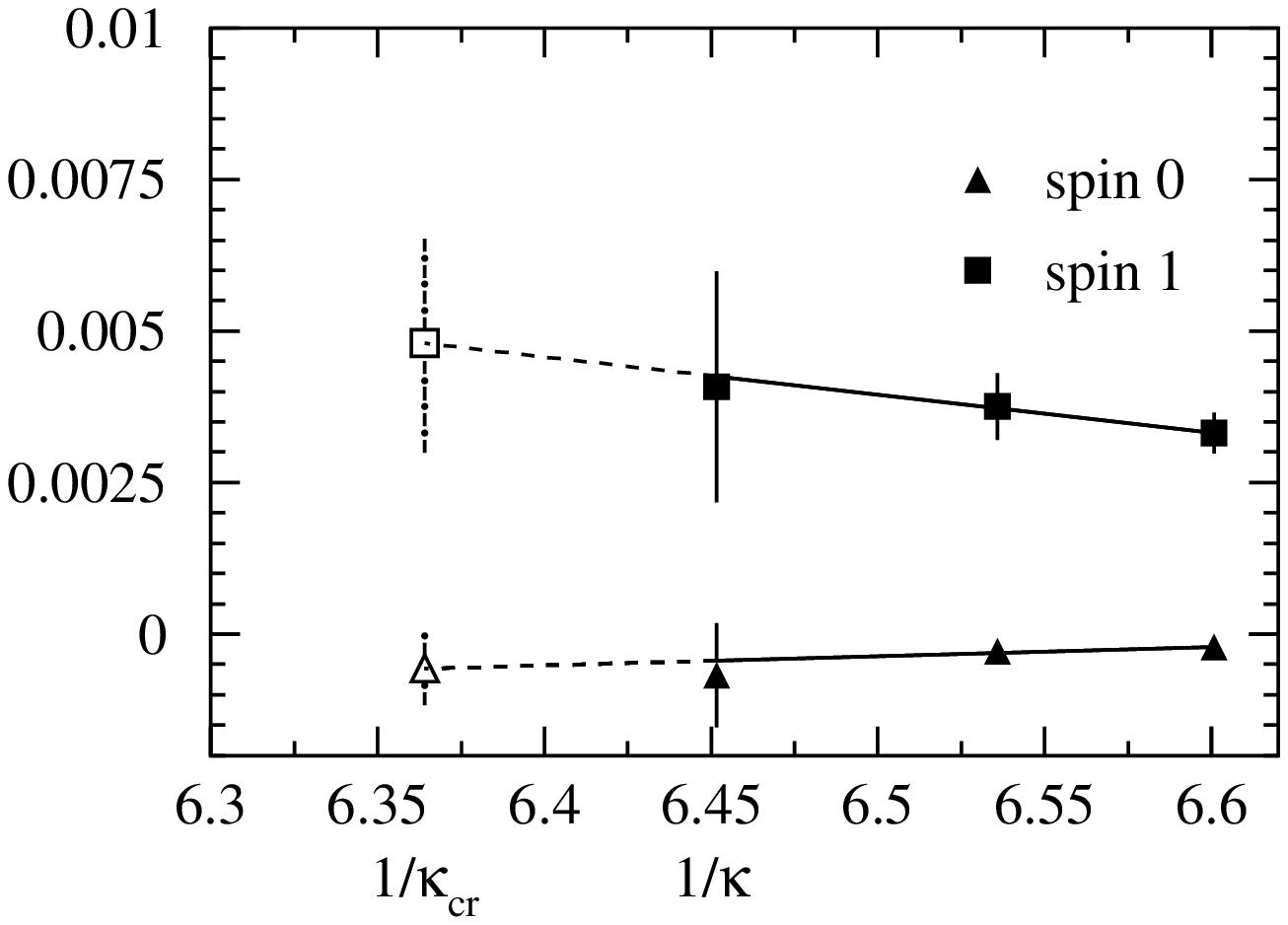,height=8cm}
\\[-1.0cm]
\epsfig{file=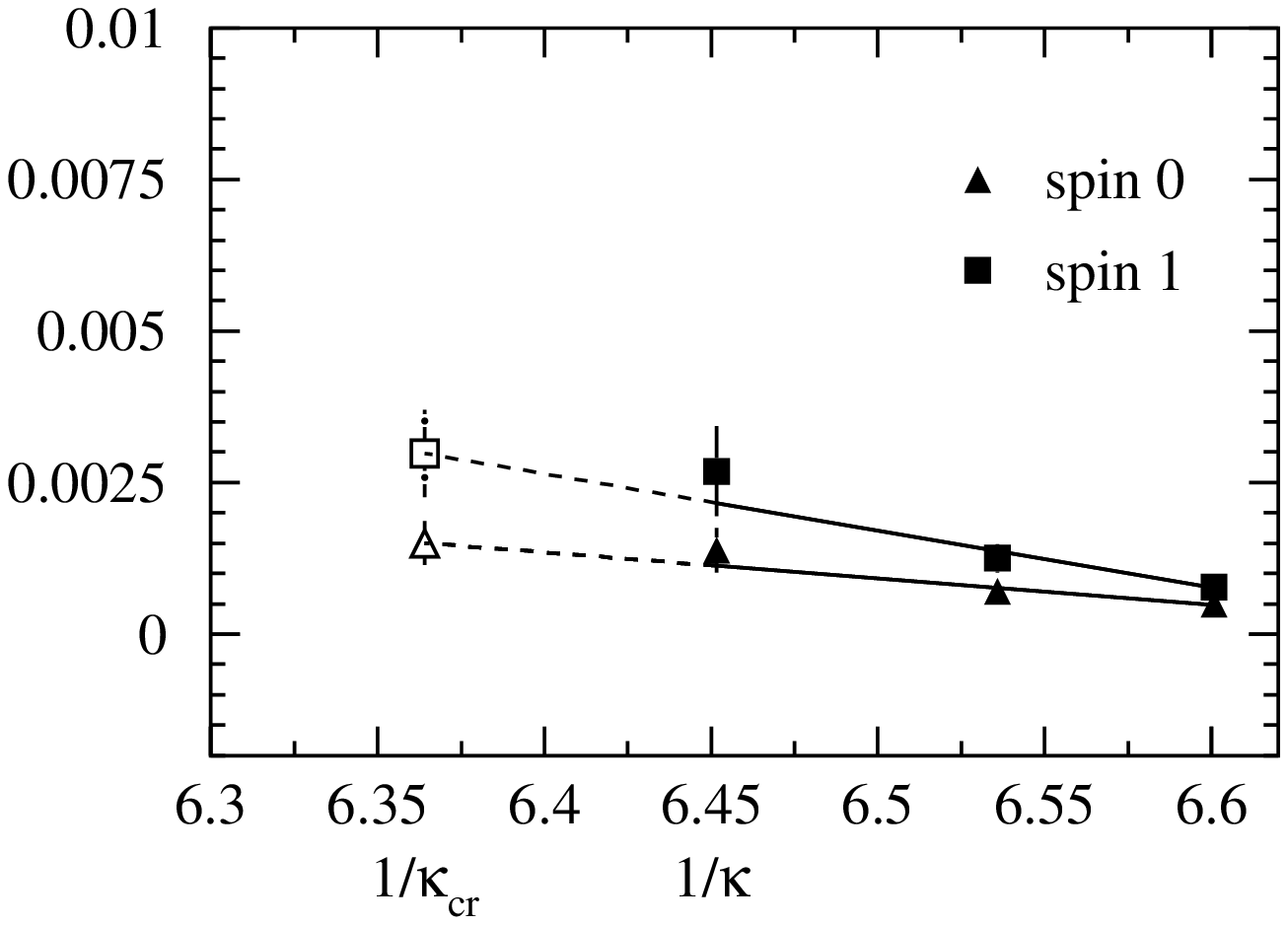,height=8cm}
\vspace*{-1.0cm}
\caption{Diquark densities with the Dirac structures
   (from top to bottom)
   $\Gamma \otimes \Gamma' = \gamma_\mu \gamma_5 \otimes \gamma_\nu \gamma_5$
   ($\overline{\bf 3}$ of colour), 
   $\sigma_{\mu \alpha} \otimes \sigma_{\nu \alpha}$
   ($\overline{\bf 3}$ of colour), 
   $\gamma_\mu \otimes \gamma_\nu $
   ($\bf 6$ of colour).}
\label{fig.dq}
\end{center}
\end{figure}

\section{DIQUARKS}

The four-quark operators can be rewritten to look like diquark densities. We 
have computed matrix elements of operators of the following form:
\begin{equation}
  \frac{1}{10}(\bar{u}_a \Gamma \gamma_5   C  \bar{u}_b^{\mathrm T})
 ( u_{a'}^{\mathrm T} C^{-1} \gamma_5 \Gamma'  u_{b'} )
 (\delta_{a b'} \delta_{b a'} -  \delta_{a a'} \delta_{b b'}) \,,
\label{diq3}
\end{equation} 
\begin{equation}
   \frac{1}{10}(\bar{u}_a \Gamma  \gamma_5  C  \bar{u}_b^{\mathrm T})
 ( u_{a'}^{\mathrm T} C^{-1} \gamma_5 \Gamma'   u_{b'} ) 
 (\delta_{a b'} \delta_{b a'} + \delta_{a a'} \delta_{b b'}) \,,
\label{diq6}
\end{equation}
where $a$, $b$, $a'$, and $b'$ are colour indices. 
The Dirac structure of the operators is represented by the matrices
$\Gamma$ and $\Gamma'$. In (\ref{diq3}) the 
diquark is in a $\overline{\bf 3}$ of colour and thus anti-symmetric in
the colour indices. In (\ref{diq6}) it is in a $\bf 6$ and symmetric 
in the colour indices.
Strictly speaking, we are again studying operators from the {\bf 27}
representation of SU(3)$_{\mathrm F}$ whose $\bar{u} \bar{u} u u$ component is
as given above. But at least within the quenched approximation the 
interpretation as valence diquark densities seems reasonable. 
In order to interpret our results we have combined the operators such that
they correspond to diquarks of spin zero and spin one.
For an operator $\cO_{\mu \nu}$ with two space-time 
indices (in Euclidean notation), e.g.\ for the Dirac structure 
$\Gamma \otimes \Gamma' = \gamma_\mu \otimes \gamma_\nu$,  
we take the expectation value of 
$\cO_{44}$ in a state with 
vanishing momentum to represent a spin-zero diquark and the expectation 
value of $\sum_{i=1}^3 \cO_{ii}$ to correspond to a spin-one diquark.

In Fig.~\ref{fig.dq} we show our results
for the diquark densities. They are plotted versus $1/\kappa$ and linearly
extrapolated to the chiral limit at $1/\kappa_{\mathrm {cr}}$, 
where the variable $1/\kappa$ determines the bare quark mass 
$m_q = (1/\kappa - 1/\kappa_{\mathrm {cr}})/(2a)$. In physical units 
these bare quark masses are approximately 100, 190, and 250 MeV.

The pattern of the results can tentatively be understood in a
non-relativistic quark picture. When the diquark is in the 
${\bf \overline{3}}$ of SU(3)$_{\mathrm c}$ it is anti-symmetric in the 
colour indices, and therefore the symmetric (in the spin indices) spin-one
state is favoured over the anti-symmetric spin-zero state. On the other
hand, when the diquark is in the (symmetric) ${\bf 6}$ of colour one might 
at first sight expect the anti-symmetric spin-zero state to dominate over 
the symmetric spin-one state. Although the spin-zero contribution is indeed
less suppressed than in the ${\bf \overline{3}}$ case it is not really
dominating. This is probably related to the fact that a diquark in the
${\bf 6}$ of SU(3)$_{\mathrm c}$ must be accompanied by (at least) one
gluon if it is to form a colour singlet with the remaining quark.
The coupling to the gluon, mixing ``large'' and ``small'' components
of the quark spinors, would invalidate the above arguments which worked
reasonably well for diquarks in the ${\bf \overline{3}}$ of colour.

\section*{ACKNOWLEDGEMENTS}
This work is supported by the DFG (Schwer\-punkt 
``Elektromagnetische Sonden'') and by BMBF. 
The numerical calculations were performed on the Quadrics 
computers at DESY Zeuthen. We wish to thank the operating staff 
for their support.

\end{document}